\documentclass[preprint,longabstract]{aastex}

\slugcomment{To appear in ApJ}

\usepackage{txfonts} \usepackage{epsfig} \usepackage{graphicx}
\usepackage{natbib} 
\begin{document}

\shorttitle{Extremely large and hot multilayer Keplerian disk around
  W51N} \shortauthors{Zapata, et al.}  \title{Extremely large and hot
  multilayer Keplerian disk around the O-type protostar W51N: The
  precursors of the HCHII regions?}

\author{Luis A. Zapata\altaffilmark{1}} \affil{Centro de
  Radioastronom\'\i a y Astrof\'\i sica, UNAM, Apdo. Postal 3-72
  (Xangari), 58089 Morelia, Michoac\'an, M\'exico} \author{Ya-Wen
  Tang} \affil{Academia Sinica Institute of Astronomy and
  Astrophysics, Taipei, Taiwan} \author{Silvia Leurini}
\affil{Max-Planck-Institut f\"{u}r Radioastronomie, Auf dem H\"ugel
  69, 53121, Bonn, Germany} \altaffiltext{1}{Max-Planck-Institut
  f\"{u}r Radioastronomie, Auf dem H\"ugel 69, 53121, Bonn, Germany}

\begin{abstract} 
We present sensitive high angular resolution (0.57$''$-0.78$''$) SO,
SO$_2$, CO, C$_2$H$_5$OH, HC$_3$N, and HCOCH$_2$OH line observations
at millimeter and submillimeter wavelengths of the young O-type
protostar W51 North made with the Submillimeter Array (SMA). We report
the presence of a large (of about 8000 AU) and hot molecular
circumstellar disk around this object, which connects the inner dusty
disk with the molecular ring or toroid reported recently, and confirms
the existence of a single bipolar outflow emanating from this object.
The molecular emission from the large disk is observed in layers with
the transitions characterized by high excitation temperatures in their
lower energy states (up to 1512 K) being concentrated closer to the
central massive protostar.  The molecular emission from those
transitions with low or moderate excitation temperatures are found in
the outermost parts of the disk and exhibits an inner cavity with an
angular size of around 0.7$''$.  We modeled all lines with a Local
Thermodynamic Equilibrium (LTE) synthetic spectra.  A detail study of
the kinematics of the molecular gas together with a LTE model of a
circumstellar disk shows that the innermost parts of the disk are also
Keplerian plus a contracting velocity.  The emission of the
HCOCH$_2$OH reveals the possible presence of a warm ``companion''
located to the northeast of the disk, however its nature is unclear.
The emission of the SO and SO$_2$ is observed in the circumstellar
disk as well as in the outflow.  We suggest that the massive protostar
W51 North appears to be in a phase before the presence of a
Hypercompact or an Ultracompact HII (HC/UCHII) region, and propose a
possible sequence on the formation of the massive stars.
\end{abstract}


\section{Introduction}

It is thought that the Hypercompact or Ultracompact HII (HCHII or
UCHII) regions trace early stages of massive star formation after the
hot molecular core phase. However, the physical processes involved in
this transition are still poorly understood \citep[see for
  reviews:][]{stan2005,liz2008}. The appearance of one HCHII begins
when the Lyman continuum output from a massive young star becomes
sufficient high to ionize its surroundings.  These objects show a
rising continuum spectra, $S_\nu \propto \nu^\alpha$, with a slope
$\alpha$ $\sim$ 1, intermediate between the optically thick and thin
limits, small sizes ($\sim$ 0.01 pc), high temperatures (10$^4$ K),
high densities ($\geq$ 10$^6$ cm$^{-3}$), and very broad radio
recombination lines ($\Delta\nu_{FWHM}$ $\geq$ 40 km s$^{-1}$)
indicating both pressure broadening and the presence of bulk motions
of dense gas \citep[e.g.][]{Sewi2004}. On the other hand, the hot
molecular cores internally heated by massive protostars are small
($\sim$ 0.05 pc) and compact regions of dense molecular gas ($\geq$
10$^6$ cm$^{-3}$) which show very rich millimeter and submillimeter
spectra \citep[see][for reviews on this
  topic]{stan2000,van2004,Cesa2005a}. The hot molecular core phase
begins at much colder temperatures ($\sim$ 100 K) than the HC/UCHII
stage.

There are many cases where the UCHII or even the HCHII regions are
observed in association with hot cores, some of them are: G5.89-0.39
\citep{su2009}; G24 A1 \citep{Bel2006}; G10.47, and G31.41+0.31
\citep{ces2010,Rol2009}; W51e2, W51e8, NGC 7538 IRS1, G28.2 and G10.6
\citep{kla2009}; IRAS 17233-3606 \citep{leu2008}.  However, there are
some others cases where the hot cores do not show the presence of a
strong HCHII \citep[e.g. AFGL490, NGC 7538S, and IRAS
  20126;][]{naka1991,sch2002,cesa2005,san2010} possibly because the
infalling molecular gas partially quenches its formation
\citep[e.g.][]{Wal1995,oso1999} or maybe because the massive central
protostar(s) are/is in an early phase with not high enough
temperatures to ionize their/its surroundings.

W51 North is one of the youngest massive stars within the luminous
cluster W51-IRS2 that is still on the process of formation. This
object is associated with a strong mm and submm source, a hot
molecular core at an excitation temperature of $\sim$ 200 K
\citep{Zhangetal1998}, no centimeter free-free emission at all ($\leq$
5 mJy at 1.3 cm) that could be related with an HC/UCHII or a thermal
jet \citep{Gaumeetal1993, Zhangetal1998, Eisneretal2002}. From this
object emanates a powerful molecular outflow observed at very small
scales by masers spots of SiO and H$_2$O, and at large scales in
thermal SiO($J=5-4$) \citep{Schnepsetal1981, Eisneretal2002,
  Imaietal2002, zapataetal2009}. Recently, \citet{zapataetal2009}
using high angular 1.3 and 0.7 mm continuum, and SO$_2$ line
observations made with the Submillimeter Array (SMA) and the Very
Large Array (VLA) resolved the large dusty and molecular hot core in a
rotating Keplerian ring with a size of $\sim$ 8000 AU \citep[at a
  distance of $\sim$ 6 kpc:] []{Imaietal2002,barba2008,xu2009}. This
structure surrounds a compact central dusty circumstellar disk with a
smaller size of approximately 3000 AU, and where the molecular
northwest(redshifted)-southeast(blueshifted) bipolar outflow
emerges. However, the relationship between both circumstellar
structures was not clear.

In this paper, we present sensitive high angular resolution SMA
submillimeter and millimeter line observations of the W51 North region
that were made in an attempt to understand the nature of the
circumstellar molecular and dusty structures associated with this
young massive protostar.  In Section 2 discuss the observations
undertook in this study. In Section 3, we present and discuss our SMA
millimeter and submillimeter data. Finally, in Section 4, we give the
main conclusions of the observations presented here.

\begin{table*}[ht]
\begin{minipage}[t]{\columnwidth}
\scriptsize \renewcommand{\footnoterule}{}
\caption{Observational and physical parameters of the submillimeter
  and millimeter lines}
\begin{center}
\begin{tabular}{lccccccc}
\hline \hline & Rest frequency & E$_{lower}$ & Range of Velocities &
Linewidth & LSR Velocity\footnote{The linewidth and LSR velocity were
  obtained fitting a Gaussian to the spectra.} & Peak Flux \\ Lines &
[GHz] & [K] & [km s$^{-1}$] & [km s$^{-1}$] & [km s$^{-1}$] & [Jy
  Beam$^{-1}$] \\ \hline \hline C$_2$H$_5$OH(37$_{8,29}$-36$_{9,28}$)
& 216.8121...& 715 & $+$54,$+$67 & 7 & 59 &
0.1\\ HCOCH$_2$OH(62$_{13,49}$-62$_{12,50}$) & 216.7585...& 1476 &
$+$54,$+$66 & 8 & 59 & 0.4\\ CO($J=3-2$)\footnote{Spectra with
  multiple peaks and in absorption about a LSR velocity of 60 km
  s$^{-1}$.} & 345.7959... & 16 & $+$20,$+$120 & 50 & 60 & 3.0
\\ SO(8$_{9}$-7$_8$)\footnote{Spectra with two peaks.}  &
346.5284... & 21 & $+$30,$+$90 & 35 & 59 &
1.0\\ SO$_2$(19$_{1,19}$-18$_{0,18}$) & 346.6521... & 151 &
$+$41,$+$72 & 15 & 58 & 1.0\\ HC$_3$N(38-37) ($\nu_t$=0) &
345.6090... & 306 & $+$51,$+$72 & 9 & 60 & 2.0
\\ HCOCH$_2$OH(68$_{19,49}$-67$_{20,48}$) & 337.3988... & 1512 &
$+$48,$+$73 & 10 & 62 & 2.0 \\ \hline \hline
\end{tabular}
\end{center}
\end{minipage}
\end{table*}

\section{Observations}

\subsection{Millimeter}

The observations were obtained with the SMA\footnote{The Submillimeter
  Array (SMA) is a joint project between the Smithsonian Astrophysical
  Observatory and the Academia Sinica Institute of Astronomy and
  Astrophysics, and is funded by the Smithsonian Institution and the
  Academia Sinica.}  during 2008 January 17. The SMA was in its very
extended configuration, which included 28 independent baselines
ranging in projected length from 30 to 385 k$\lambda$.  The phase
reference center of the observations was R.A. = 19h23m40.05s, decl.=
14$^\circ$31$'$05.0$''$ (J2000.0).  The frequency was centered at
217.1049 GHz in the Lower Sideband (LSB), while the Upper Sideband
(USB) was centered at 228.1049 GHz. The primary beam of the SMA at 230
GHz has a FWHM of about 60$''$.

We detected the lines C$_2$H$_5$OH(37$_{8,29}$-36$_{9,28}$) and
HCOCH$_2$OH(62$_{13,49}$-62$_{12,50}$) in the LSB. See Table 1 for
their rest frequencies and rotational temperatures above of the ground
state. The full bandwidth of the SMA correlator was 4 GHz (2 GHz in
each band). The SMA digital correlator was configured in 24 spectral
windows (``chunks'') of 104 MHz each, with 256 channels distributed
over each spectral window, providing a resolution of 0.40 MHz (0.56 km
s$^{-1}$) per channel. However, in this study we smoothed our spectral
resolution to about 1 km s$^{-1}$.

The zenith opacity ($\tau_{230 GHz}$), measured with the NRAO tipping
radiometer located at the Caltech Submillimeter Observatory, was
$\sim$ 0.15, indicating good weather conditions during the
observations.  Observations of Uranus provided the absolute scale for
the flux density calibration.  Phase and amplitude calibrators were
the quasars 1925+211 and 2035+109.  The uncertainty in the flux scale
is estimated to be 15--20$\%$, based on the SMA monitoring of quasars.
Observations of Uranus provided the absolute scale for the flux
density calibration.  Further technical descriptions of the SMA and
its calibration schemes can found in \citet{Hoetal2004}.

The data were calibrated using the IDL superset MIR, originally
developed for the Owens Valley Radio Observatory
\citep{Scovilleetal1993} and adapted for the SMA.\footnote{The MIR-IDL
  cookbook by C. Qi can be found at
  http://cfa-www.harvard.edu/$\sim$cqi/mircook.html} The calibrated
data were imaged and analyzed in the standard manner using the MIRIAD
and AIPS packages.  We used the ROBUST parameter of the INVERT task of
MIRIAD set to 2 to obtain a slightly better sensitivity while losing
some angular resolution. The resulting image rms noise of line images
was around 30 mJy beam$^{-1}$ for each velocity channel (with a
smoothed size of 1 km s$^{-1}$) at an angular resolution of
$0\rlap.{''}57$ $\times$ $0\rlap.{''}42$ with a P.A. = $57.6^\circ$.

\begin{figure}[ht]
\begin{center}
\includegraphics[scale=0.4]{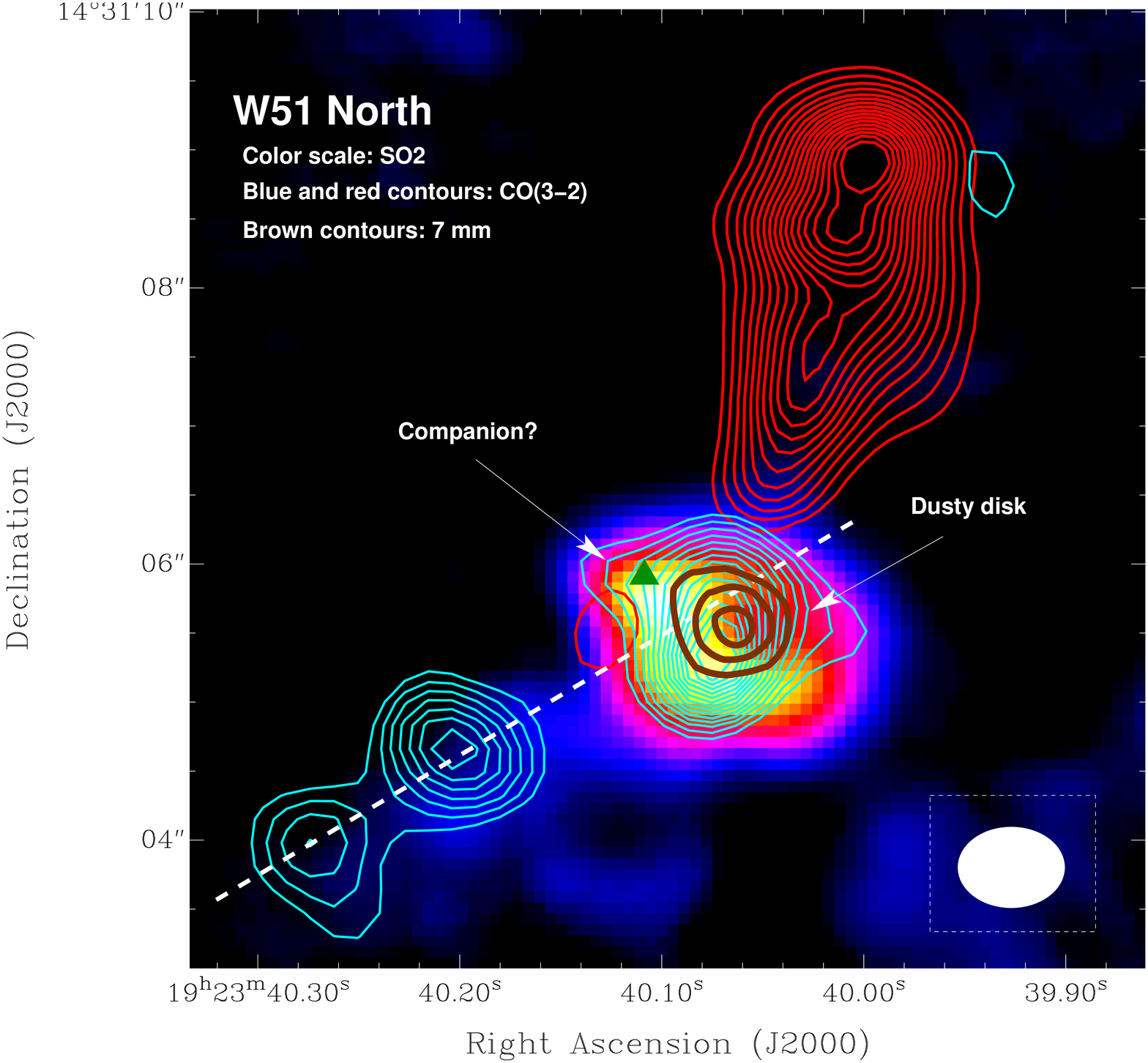}
\caption{\scriptsize Overlay of the SO$_2$(22$_{2,20}$-22$_{1,21}$)
  moment zero emission (color image), the 7 mm continuum emission from
  the compact dusty disk (brown contours), and the CO($J=3-2$) moment
  zero emission from the bipolar outflow (blue and red contours). The
  SO$_2$ and the 7 mm images were taken from
  \citet{zapataetal2009}. The integrated velocity range for the
  CO($J=3-2$) is from +20 to +50 km s$^{-1}$ for the blueshifted gas,
  and from +70 to +120 km s$^{-1}$ for the redshifted gas. The brown
  contours are the 10\%, 13\%, and 16\% of the peak of the 7 mm
  continuum emission, the peak is 100 mJy beam$^{-1}$.  The
  synthesized beam of the CO($J=3-2$) contour image is 0.71$''$
  $\times$ 0.57$''$ with a P.A. of 87.6$^\circ$ and is shown in the
  bottom right corner. The green triangle marks the position of the
  possible companion reported in this paper. The white dashed line
  shows how the blueshifted side of the outflow seems to be more
  likely energized from the central massive protostar W51 North rather
  than the possible companion.}
\label{fig1}
\end{center}
\end{figure}

\begin{figure*}[ht]
\begin{center}
\includegraphics[scale=0.31]{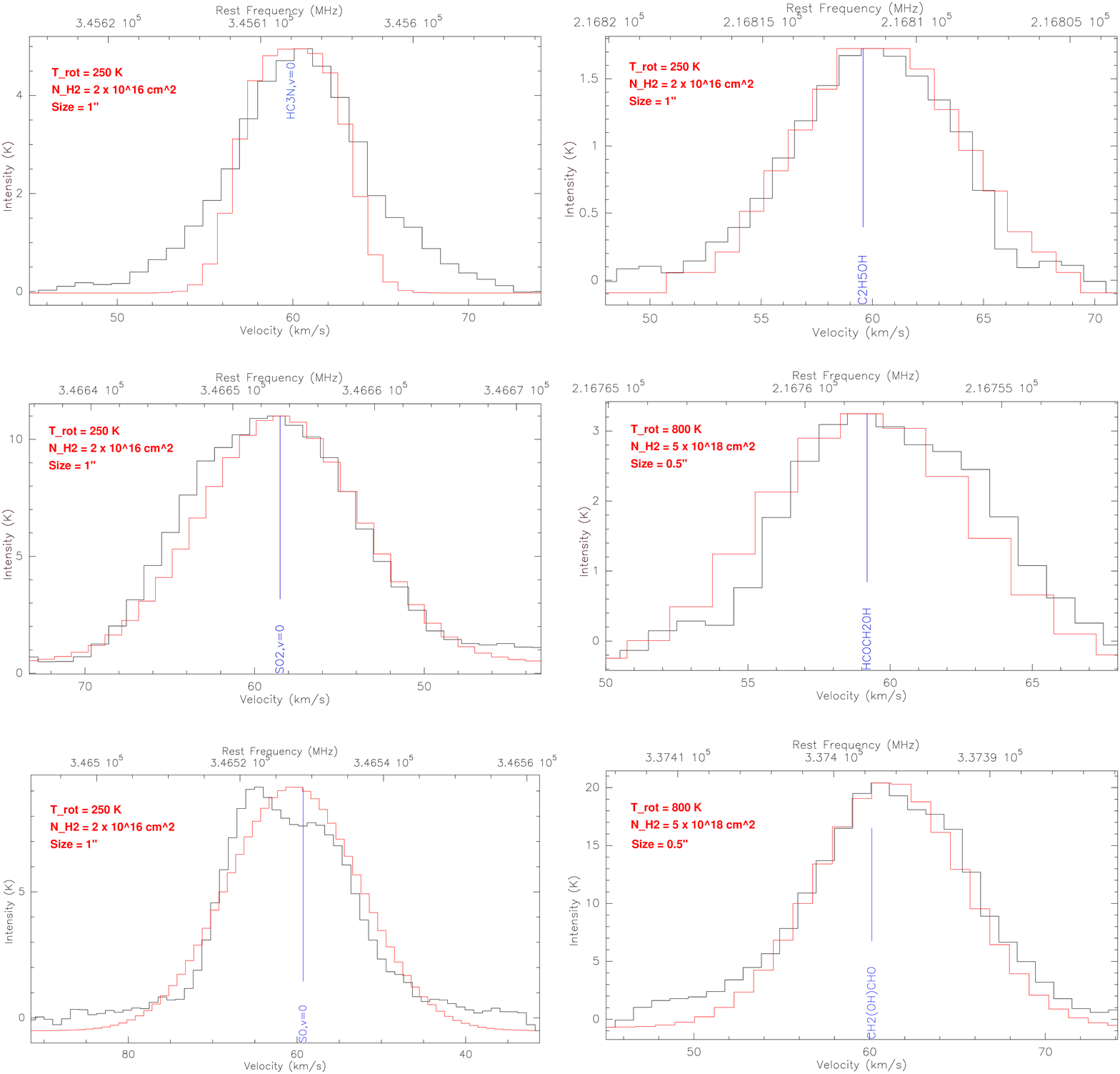}
\caption{\scriptsize Spectra from the selected millimeter and
  submillimeter lines observed toward W51 North. The red line
  represents in all panels the synthetic spectrum obtained for the
  best fitting solution with the parameters shown in each panel.  The
  spectra was obtained from the average of the total emission in each
  line using the task ``imspect'' of MIRIAD.}
\label{fig2}
\end{center}
\end{figure*}

\begin{figure*}[ht]
\begin{center}
\includegraphics[scale=0.20]{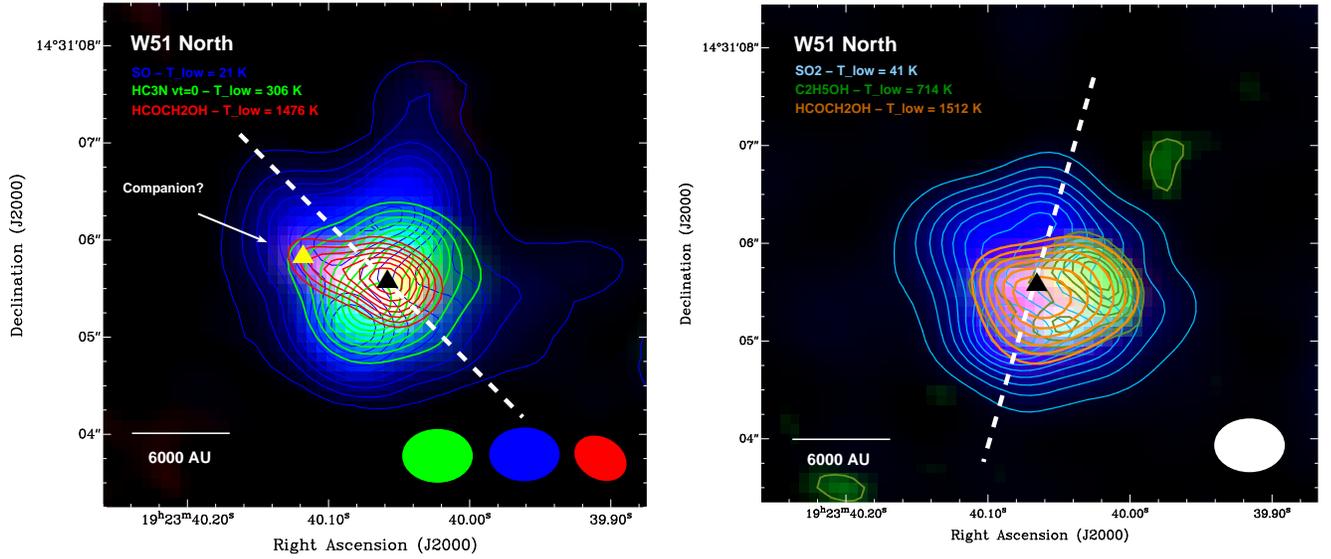}
\caption{\scriptsize Composite images of the molecular emission from
  the circumstellar large disk around the massive protostar W51 North.
  Left: The blue color scale and the contours are representing the
  integrated emission of the SO(8$_{9}$-7$_8$), green ones the
  HC$_3$N(38-37) ($\nu_t$=0), and the red ones the
  HCOCH$_2$OH(62$_{13,49}$-62$_{12,50}$).  The blue contours are from
  5\% to 90\% with steps of 10\% of the peak of the line emission, the
  peak is 53 Jy beam$^{-1}$ km s$^{-1}$.  The green contours are from
  30\% to 90\% with steps of 10\% of the peak of the line emission,
  the peak is 26 Jy beam$^{-1}$ km s$^{-1}$.  The red contours are
  from 30\% to 90\% with steps of 10\% of the peak of the line
  emission, the peak is 3.5 Jy beam$^{-1}$ km s$^{-1}$.  The
  synthesized beams of the millimeter and submillimeter observations
  are shown on the right bottom corner. Right: The light blue color
  scale and the contours are representing the integrated emission of
  the SO$_2$(19$_{1,19}$-18$_{0,18}$), green ones the
  C$_2$H$_5$OH(37$_{8,29}$-36$_{9,28}$), and the orange ones the
  HCOCH$_2$OH(68$_{19,49}$-67$_{20,48}$).  The light blue contours are
  from 30\% to 90\% with steps of 10\% of the peak of the line
  emission, the peak is 38 Jy beam$^{-1}$ km s$^{-1}$.  The green
  contours are from 30\% to 90\% with steps of 10\% of the peak of the
  line emission, the peak is 2 Jy beam$^{-1}$ km s$^{-1}$.  The orange
  contours are from 30\% to 90\% with steps of 10\% of the peak of the
  line emission, the peak is 27 Jy beam$^{-1}$ km s$^{-1}$.  The
  synthesized beam of the three measurements is the same and is shown
  on the right bottom corner.  The triangles represent the position of
  the two possible embedded massive protostars as suggested by the hot
  molecular emission of the HCOCH$_2$OH(62$_{13,49}$-62$_{12,50}$) and
  HC$_3$N(38-37) ($\nu_t$=0). The white dashed line in both panels
  represents the position and orientation where the PV-diagrams shown
  in Figures 4 and 5 were computed. }
\label{fig3}
\end{center}
\end{figure*}

\begin{table*}[ht]
\begin{minipage}[t]{\columnwidth}
\scriptsize \renewcommand{\footnoterule}{}
\caption{Observational parameters of the circumstellar multilayer
  disk}
\begin{center}
\begin{tabular}{lccccccc}
\hline \hline & \multicolumn{2}{c}{Central Position} & Flux Density &
\multicolumn{2}{c}{Deconvolved Size\footnote{The Gaussian fitting was
    obtained with the task JMFIT of AIPS.}} & Approx. cavity
size\\ \hline & $\alpha(2000)$ & $\delta(2000)$ & & & & \\ Specie & [h
  m s] & [$\circ$ $''$ $'$] & [Jy Beam$^{-1}$ km s$^{-1}$] &
[arcsec$^2$]& [Degrees] & [arcsec$^2$] \\ \hline \hline
SO(8$_{9}$-7$_8$) & 19 23 40.074 & 14 31 05.70 & 341$\pm$20 &
1.78$\pm$0.01 $\times$ 1.39$\pm$0.02 & 171$\pm$2 & $\sim$
0.8\\ SO$_2$(19$_{1,19}$-18$_{0,18}$) & 19 23 40.071 & 14 31 05.54 &
234$\pm$15 & 1.65$\pm$0.04 $\times$1.49$\pm$0.03 & 24$\pm$6& $\sim$
0.7\\ HC$_3$N(38-37) ($\nu_t$=0) & 19 23 40.060 & 14 31 05.54 &
88$\pm$15 & 1.36$\pm$0.04 $\times$0.93$\pm$0.04 &135$\pm$7 & $\sim$
0.6\\ C$_2$H$_5$OH(37$_{8,29}$-36$_{9,28}$) & 19 23 40.041 & 14 31
05.49 & 116$\pm$15 & 1.31$\pm$0.03 $\times$0.89$\pm$0.03 & 107$\pm$7&
$\sim$ 0.8\\ HCOCH$_2$OH(68$_{19,49}$-67$_{20,48}$) & 19 23 40.054 &
14 31 05.44 & 105$\pm$13 & 1.14$\pm$0.03 $\times$0.88$\pm$0.03 &
99$\pm$3& --\\ HCOCH$_2$OH(62$_{13,49}$-62$_{12,50}$) &19 23 40.062 &
14 31 05.59 & 9.1$\pm$2 & 0.81$\pm$0.03 $\times$ 0.40$\pm$0.09 &
75$\pm$3 & --\\ ---Companion & 19 23 40.077 & 14 31 05.63 & 8.5$\pm$1
& 1.27$\pm$0.04 $\times$0.39$\pm$0.03 & 74$\pm$2& --\\ \hline & & &
Dusty circumstellar disk\footnote{Data obtained from
  \citet{zapataetal2009}. } & & & \\ 7 mm & 19 23 40.057 & 14 31 05.67
& -- & 0.58$\pm$0.02 $\times$0.27$\pm$0.02 & 70$\pm$6&
--\\ \hline\hline
\end{tabular}
\end{center}
\end{minipage}
\end{table*}

\subsection{Submillimeter}

The observations were obtained with the SMA on 2008 July 13.  At the
time of these observations the SMA had seven antennas in its extended
configuration with baselines ranging in projected length from 30 to
255 k$\lambda$.  The primary beam of the SMA at 340 GHz has a FWHM of
37$''$. The molecular emission from the hot core was found well inside
of the primary beam.

The receivers were tuned to a frequency of 345.796 GHz in the upper
sideband (USB), while the lower sideband (LSB) was centered on 335.796
GHz.  The LSB contained line HCOCH$_2$OH(68$_{19,49}$-67$_{20,48}$)
while the USB the lines SO(8$_{9}$-7$_8$),
SO$_2$(19$_{1,19}$-18$_{0,18}$), CO($J=3-2$), and HC$_3$N(38-37)
($\nu_t$=0). See Table 1 for their rest frequencies and lower level
energies above the ground energy state.  The SMA digital correlator
was configured in 24 spectral windows (``chunks'') of 104 MHz each,
with 128 channels distributed over each spectral window, thus
providing a spectral resolution of 0.81 MHz (0.70 km s$^{-1}$) per
channel.  However, we smoothed our spectral resolution to 1.0 km
s$^{-1}$ per spectral channel.

The zenith opacity ($\tau_{230 GHz}$) was $\sim$ 0.035 -- 0.04,
indicating excellent weather conditions.  Observations of titan
provided the absolute scale for the flux density calibration.  Phase
and amplitude calibrators were the quasars 1751$+$096, and 1925$+$211.
The uncertainty in the flux scale is also estimated to be 15 --
20$\%$, also based on the SMA monitoring of the quasars.

The calibrated data were imaged and analyzed in standard manner using
the MIRIAD, and AIPS packages. We also used the ROBUST parameter set
to 2.  The line image rms noise was around 100 mJy beam$^{-1}$ for
each velocity channel (with a smoothed size of 1 km s$^{-1}$) at an
angular resolution of $0\rlap.{''}78$ $\times$ $0\rlap.{''}58$ with a
P.A. = $84.7^\circ$.

\section{Results}

In Figure 1, we present the disk/ring/outflow system found towards W51
North using observations of the SiO($J=5-4$),
SO$_2$(22$_{2,20}$-22$_{1,21}$), and millimeter continuum by
\citet{zapataetal2009}. However, in this image, we have exchanged the
SiO emission by the emission of the CO($J=3-2$) presented here for the
first time. The CO($J=3-2$) confirms the presence of a single powerful
and collimated (15$^\circ$) bipolar outflow (with the blueshifted side
towards the southeast while the redshifted one to the north) emanating
from the disk/ring structure at similar scales. However, the
CO($J=3-2$) reveals that the outflow extents for a few arcseconds more
on both sides in comparison with the emission traced by SiO($J=5-4$),
and shows a slightly larger radial velocity range ($+$30 to $+$120 km
s$^{-1}$).  In this image, it is more clear to see the deviation of
the blueshifted side of the outflow with respect to the redshifted one
than in \citet{zapataetal2009}.

Figure 2 shows the spectra of the multiple lines reported here with
exception of the CO($J=3-2$).  Additionally, we present synthetic LTE
models for the spectra using XCLASS program
\citep[][]{Comitoetal2005}.  The data were fitted well with the
synthetic model using rotational temperatures, column densities, and
sizes of 250 K, 2 $\times$ 10$^{16}$ cm$^2$, and 1$''$, respectively,
for the lines SO(8$_{9}$-7$_8$), SO$_2$(19$_{1,19}$-18$_{0,18}$),
HC$_3$N(38-37) ($\nu_t$=0), and C$_2$H$_5$OH(37$_{8,29}$-36$_{9,28}$)
and 800 K, 5 $\times$ 10$^{18}$ cm$^2$, and 0.5$''$ for transitions
with high excitation temperatures in the lower energy states,
HCOCH$_2$OH(68$_{19,49}$-67$_{20,48}$) and
HCOCH$_2$OH(62$_{13,49}$-62$_{12,50}$).  This modeling serves as for
avoiding the line confusion, and suggest that hot gas is only found in
the innermost parts of the hot core.  See Table 1 for the rest
frequencies and the physical parameters of the lines. The physical
parameters of all lines were obtained fitting a Gaussian to the line
profile.

In order to be more confident about the detection of the
glycol-aldehyde (HCOCH$_2$OH) toward W51 North, we have fitted
simultaneously many other lines that fall in our 2 GHz bands with our
synthetic LTE model using the same parameters as above. We obtained a
reasonable agreement between the synthetic model and the detected
lines.  A full astrochemical analysis of these lines (and other
species) is beyond the scope of the present paper in which we
concentrate on the spatial distributions of the lines presented in
Table 1.

The spectra of the SO(8$_{9}$-7$_8$) additionally shows
self-absorption possibly due to the molecular gas is falling into the
central object \citep[][]{Zapataetal2008}.

\begin{figure*}[ht]
\begin{center}
\includegraphics[scale=0.37, angle=0]{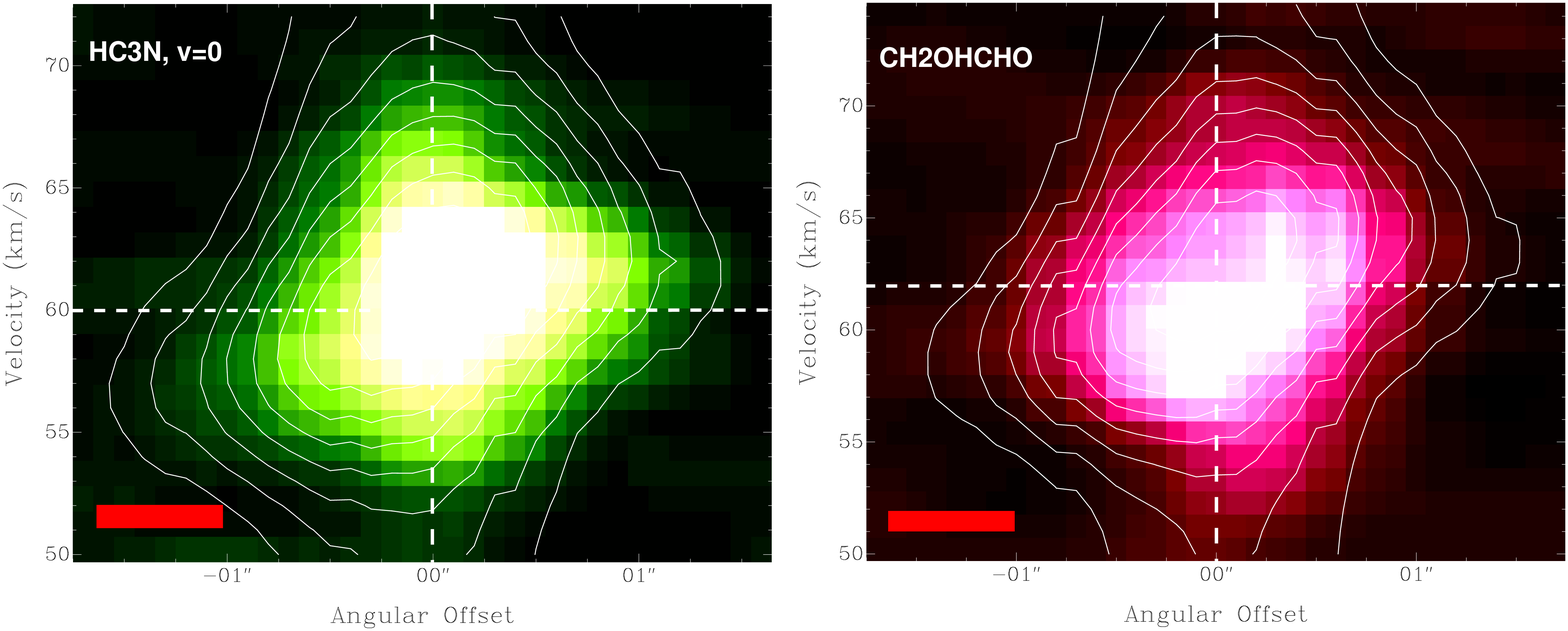}
\caption{\scriptsize Position-velocity diagrams of the HC$_3$N(38-37)
  (green scale color), and the HCOCH$_2$OH(68$_{19,49}$-67$_{20,48}$)
  (pink scale color) line emission from the circumstellar disk. The PV
  diagrams were computed at a P.A. = 30$^\circ$ (see Figure 3). These
  diagrams are additionally overlaid with the position-velocity
  diagram at same P.A. of our Keplerian disk model (contours). The
  contours are from 20\% to 90\% with steps of 10\% of the peak of the
  line emission of our model. The units of the horizontal axis are in
  arcseconds. The systemic LSR radial velocity of the ambient
  molecular cloud is about 60 km s$^{-1}$, see Table 1. The angular
  and spectral resolutions are shown in each panel on the bottom left
  corner.  The synthesized beam of both images is 0.71$''$ $\times$
  0.57$''$ with a P.A. of 87.6$^\circ$. The spectral resolution was
  smoothed to 1 km s$^{-1}$.  The origin in the horizontal axis is at
  R.A. = 19h23m40.05s, decl.= 14$^\circ$31$'$05.0$''$ (J2000.0).}
\label{fig4}
\end{center}
\end{figure*}

\begin{figure*}[ht]
\begin{center}
\includegraphics[scale=0.4]{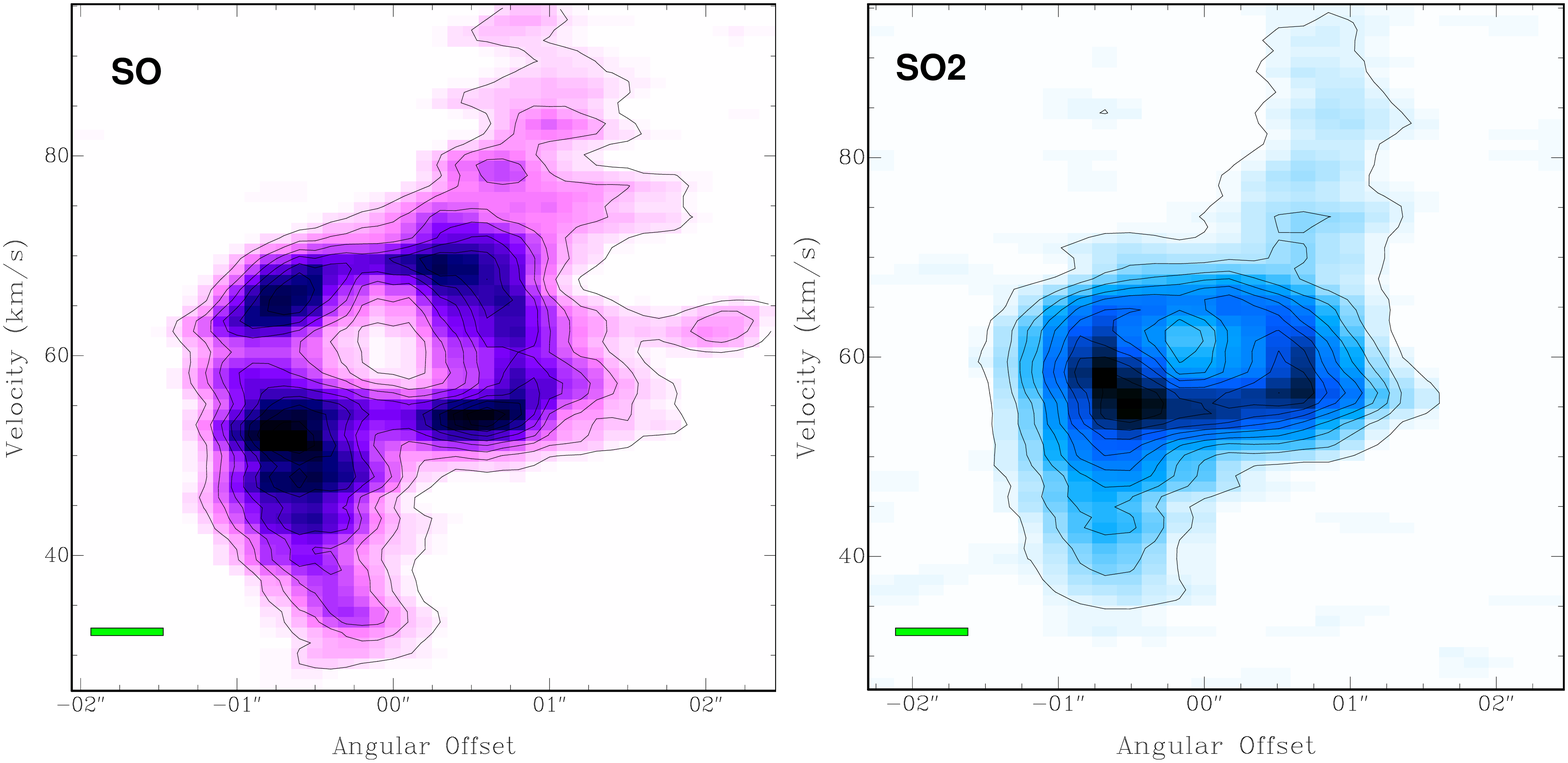}\\
\caption{\scriptsize Position-velocity diagrams for the
  SO(8$_{9}$-7$_8$) and the SO$_2$(19$_{1,19}$-18$_{0,18}$) from the
  large circumstellar disk and the outflow.  They were computed at a
  P.A. of 160$^\circ$ (see Figure 3). The contours in both panels are
  from 10\% to 90\% with steps of 10\% of the peak of the line
  emission. The units of the horizontal axis are in arcseconds. The
  systemic LSR radial velocity of the ambient molecular cloud is about
  60 km s$^{-1}$, see Table 1. The angular and spectral resolutions
  are shown in each panel on the bottom left corner.  The synthesized
  beam of both images is 0.71$''$ $\times$ 0.57$''$ with a P.A. of
  87.6$^\circ$. The spectral resolution was smoothed to 1 km
  s$^{-1}$. The origin in the horizontal axis is at R.A. =
  19h23m40.05s, decl.=14$^\circ$31$'$05.0$''$ (J2000.0).}
\end{center}
\label{fig5}
\end{figure*}

The integrated intensity or moment zero maps of the lines associated
with the molecular disk are presented in Figure 3. We have divided
these lines into three groups showing the transitions with energy
levels with moderate (20-160 K), high (300-800 K), and very high
(1400-1600 K) energies and presented them in two different panels.
From these images, it is easy to see how the the molecular emission is
found in layers with the transitions characterized by high excitation
temperatures in their lower energy states (up to 1512 K), and/or
critical densities, being concentrated closer to the central high-mass
protostar W51 North.  This phenomenon confirms clearly that the
heating in W51 North is internal.  Additionally, molecules with
transitions characterized by low and moderate excitation temperatures
{\it i.e.} SO(8$_{9}$-7$_8$), SO$_2$(19$_{1,19}$-18$_{0,18}$),
HC$_3$N(38-37) ($\nu_t$=0), and C$_2$H$_5$OH(37$_{8,29}$-36$_{9,28}$)
show a central cavity of approximately an angular size of 0.7$''$, see
Table 2. These lines trace thus the molecular Keplerian ring (or
toroid) around W51 North reported in SO$_2$ by
\citet[][]{zapataetal2009}, see Figure 1.  The transition of the
molecule SO$_2$ utilized by these authors has a similar excitation
temperature in the lower energy state as those mentioned above. Note
the similarity between the angular size of the inner cavity and the
size of the compact dusty circumstellar disk reported by
\citet[][]{zapataetal2009}, see Table 2.

The molecular emission from the transition with a high excitation
temperature in the lower energy states ({\it i.e.}
HCOCH$_2$OH(68$_{19,49}$-67$_{20,48}$)) is found peaking at the
position of the cavity and showing similar angular sizes to this, see
Table 2. This molecule is thus more intimately related with the
circumstellar compact dusty disk.

The observational parameters of all lines are shown in Table 2.  We
noted from Table 1 and 2, there is a correlation between the
deconvolved sizes of the molecular emission and the excitation
temperatures in lower energy states of each molecular specie. The
molecular emission from transitions characterized by high excitation
temperatures show to be very compact.  This correlation could be
obviously obtained for lines that are optically thick with T$_b$
$\sim$ $\eta$ T$_{ex}$, where the T$_b$ is the brightness temperature,
$\eta$ is the filling factor, and T$_{ex}$ is the excitation
temperature.

In the left panel of Figure 3 the emission of the lines
HCOCH$_2$OH(62$_{13,49}$-62$_{12,50}$) and HC$_3$N(38-37) ($\nu_t$=0)
reveals the presence of a possible warm ``companion'' located to the
northeast of the disk. We have marked the position of this putative
companion with a yellow triangle. This ``companion'' is also observed
in SO$_2$ at the same position, see Figure 3 of
\citet[][]{zapataetal2009}, and is associated with a group of water
maser spots \citep{Imaietal2002,Eisneretal2002}. The position of the
``companion'' was found by fitting a Guassian to its
HCOCH$_2$OH(62$_{13,49}$-62$_{12,50}$)) emission. However, it still
not clear if the ``companion'' is real protostar or if this could be
the result of the interaction of the outflow with a high density zone
of the molecular cloud.

We do not find any clear evidence of outflowing gas activity
associated with this possible ``companion'' \citep[see
  also][]{zapataetal2009}. Although the blueshifted side of the
CO($J=3-2$) outflow has not the same position angle as the redshifted
one, this seems not be ejected from the companion.  We show how the
blueshifted side appears to be more likely ejected from W51 North in
Figure 1. We drew a line that crosses this side of the outflow and
points directly to the dusty compact disk.  The SO$_2$ and SiO
emission indeed show how the outflow is deviated to where the
CO($J=3-2$) is located \citep[][]{zapataetal2009}.  However, more
observations in some other molecular outflow tracers are thus
necessity to firmly discard the existence of a second outflow in W51
North. It is interesting to note that the possible companion is not
observed in the hotter molecular gas tracer
HCOCH$_2$OH(68$_{19,49}$-67$_{20,48}$) suggesting that it may not as
warm as the central massive star.

Figures 4 and 5 show the position-velocity diagrams (PV-diagrams) of
different molecules which trace distinct scales of the disk. In Figure
3, we have marked the orientation and position of the PV-cuts. In the
left panel of Figure 3 we show a white line with a P.A.=30$^\circ$
that corresponds to the PV-cuts shown in Figure 4. On the other hand,
the white line with a P.A.=160$^\circ$ in the right panel corresponds
to the the PV-cuts shown in Figure 5.

Figure 4 shows the molecular emission from HC$_3$N(38-37) and
HCOCH$_2$OH(68$_{19,49}$-67$_{20,48}$) located in the innermost parts
of the disk, while the PV-diagrams of molecules as SO(8$_{9}$-7$_8$),
and SO$_2$(19$_{1,19}$-18$_{0,18}$), which trace the outermost parts,
are presented in Figure 5. The PV-diagrams in Figure 4 reveal that the
hot molecular gas closer to the massive protostar is Keplerian.  In
addition, in this figure we have overlaid the PV-diagram of the LTE
Keplerian disk modeled in \citet[][]{zapataetal2009}, but without an
inner cavity and a smaller size. Both structures shown a very good
correspondence.  The SO(8$_{9}$-7$_8$) and
SO$_2$(19$_{1,19}$-18$_{0,18}$) presented in Figure 5 trace much
larger structures similar to the ring reported in
\citet[][]{zapataetal2009}. These molecules in addition show clearly
two northwest and southeast high velocity extensions excited by the
bipolar outflow mapped in CO($J=3-2$) and SiO($J=5-4$).  The
PV-diagram of the molecule C$_2$H$_5$OH(37$_{8,29}$-36$_{9,28}$) shows
also Keplerian motions, while that obtained from the molecule
HCOCH$_2$OH(62$_{13,49}$-62$_{12,50}$) shows a more compact structure
without a clear morphology, these diagrams are not presented in this
study.

Finally, in Figure 6 we show the spectral energy distribution (SED)
from the centimeter, millimeter, and submillimeter wavelengths of W51
North with data obtained from \citet[][]{zapataetal2009} and Tang et
al. (in prep.). The new value at submillimeter wavelengths presented
here shows also to be good fitted with the $\alpha$ = 2.8 value (where
the flux density goes as S$_\nu$=$\nu^\alpha$) found by
\citet[][]{zapataetal2009}.

\begin{figure}[h!]
\begin{center}
\includegraphics[scale=0.6]{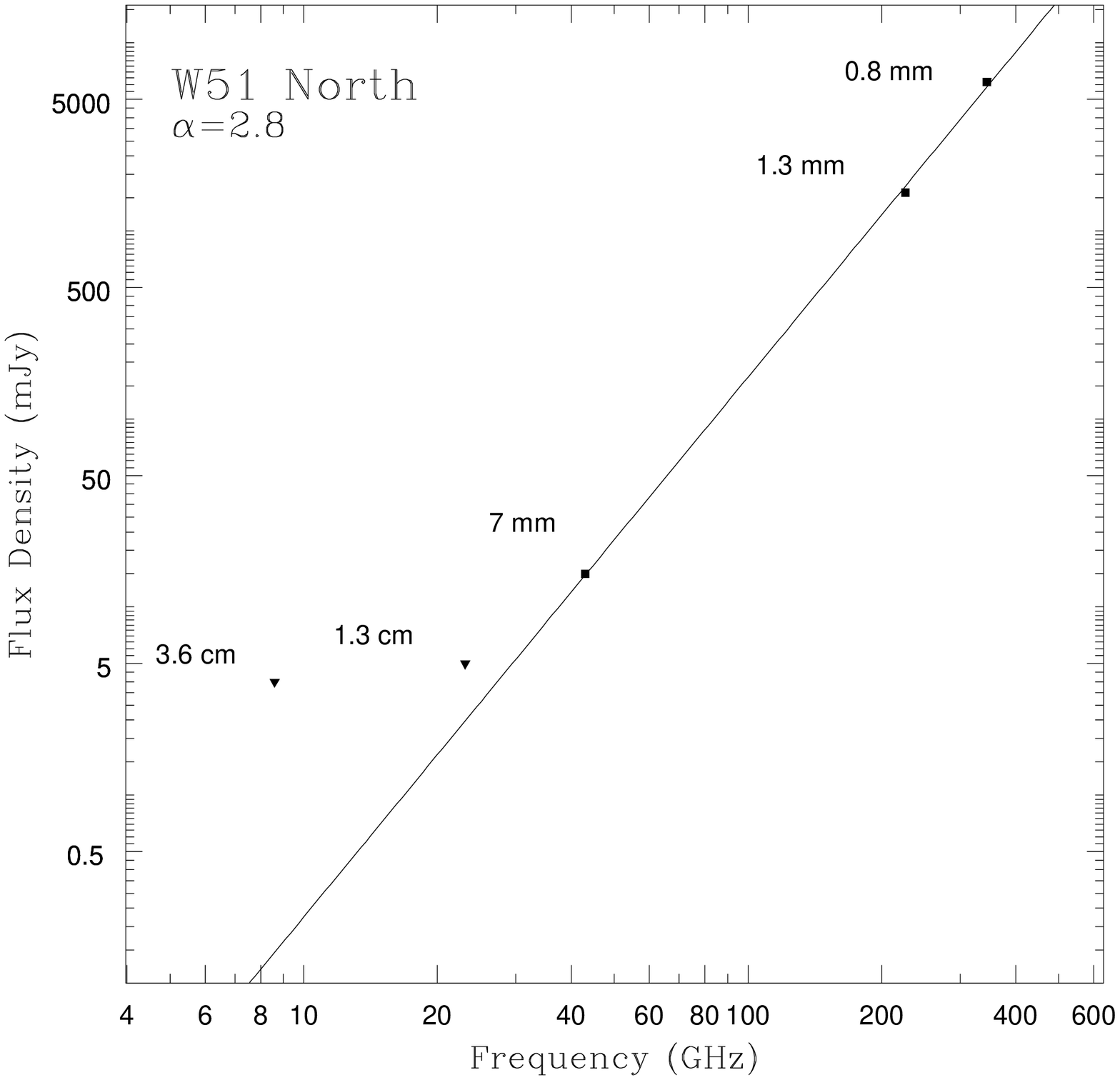}\\
\caption{\scriptsize SED for the source W51 North combining 3.6 cm,
  1.3 cm, 7 mm, 1.3 mm and 0.8 mm continuum data. The millimeter data
  were obtained from \citet{zapataetal2009} and the submillimeter data
  was obtained from Tang et al. (in prep.). The respective error bars
  were smaller than the squares and are not presented here. The line
  is a least-squares power-law fit (of the form S$_\nu$ $\propto$
  $\nu^\alpha$) to the spectrum. Note that the value of $\alpha$=2.8
  obtained by \citet{zapataetal2009} fit very well all millimeter and
  submillimeter measurements. The 1.3 cm point is a upper limit
  (5$\sigma$=5 mJy) obtained from Figure 1 of
  \citet{Eisneretal2002}. The 3.6 cm point is a upper limit
  (5$\sigma$=4 mJy) obtained from Figure 5 of \citet{meh1994}.}
\end{center}
\label{fig6}
\end{figure}

\begin{figure*}[ht]
\begin{center}
\includegraphics[scale=0.75, angle=0]{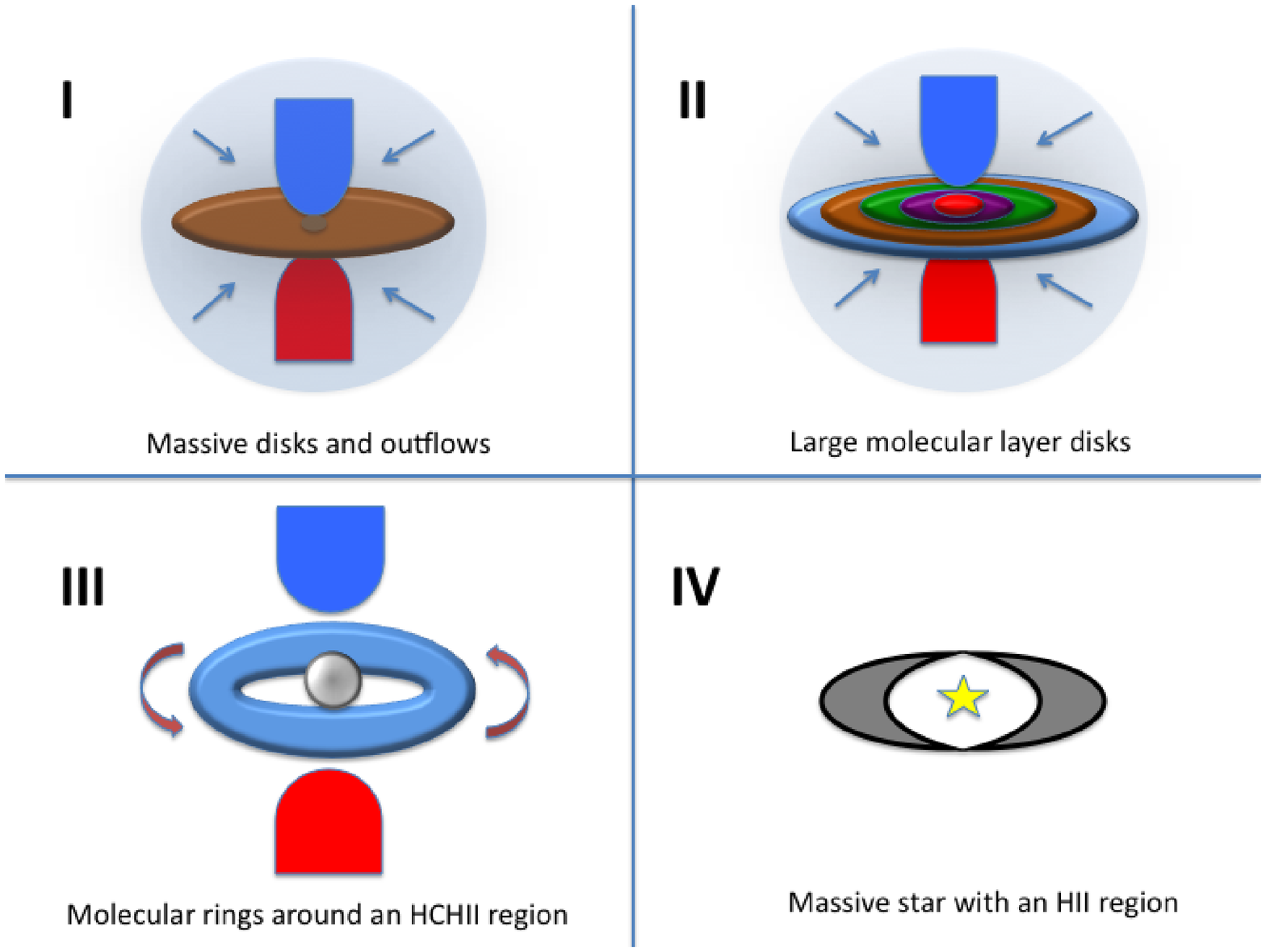}\\
\caption{\scriptsize A very simple diagram showing the different
  evolutionary phases of a massive protostar. In this diagram, we do
  not included multiplicity for simplicity. }
\end{center}
\label{fig7}
\end{figure*}

\section{Discussion}

\subsection{A multilayer infalling and keplerian disk around W51 North:
Connecting the molecular ring and dusty compact disk.}

The combined millimeter and submillimeter line observations presented
here from the ring and compact dusty disk structures reported in
\citet{zapataetal2009} suggest that we are seeing a single hot
multilayer accreting disk around W51 North (Figure 3). This large
molecular disk connects both circumstellar structures, the dusty
compact disk, observed at 7 and 1 mm, and the molecular ring mapped in
SO$_2$ in a single structure.  Some molecules are found in association
with the dusty compact disk while some others with the ring.

The molecular emission associated with the multilayer disk reveals
that the molecular emission with transitions with a low excitation
temperature in the lower energy states delineates the outer disk and
shows a cavity, while such transitions with a high excitation
temperature in the lower energy state are found only in the innermost
parts of the disk and closer to the central massive protostar(s).  One
would expect to this physical phenomenon happen in disks or cores
around high- and low-mass protostar(s) due to the large optical depths
and higher temperatures only found close to the young star. Up to date
detailed models of the ensuing chemical evolution at these much higher
temperatures do not at present exist.

It still is not clear why we do see an inner cavity on the disk in
some molecules with transitions characterized by low and moderate
excitation temperatures.  This cavity maybe is not a real hole on the
disk rather this could be due to photo-dissociation or opacity effects
of these molecules toward the innermost parts of the disk.  However,
we do not discard the possibility that the cavity could be real and
formed possibly by the tidal effects of the young multiple massive
protostars in the middle of the ``ring'' as in the case of Ori 139-409
located in Orion South where the protostars have lower masses
\citep{zap2010}.  The molecular emission with transitions
characterized by high excitation temperatures in the lower energy
states may arise from the possible compact circumstellar disks inside
of the cavity that with the actual angular resolution are not
resolved.  The presence of a binary system at the center of W51 North
has been suggested by the precession of the molecular outflow at very
small scales found by \citet{Eisneretal2002}. This precession
phenomenon may also explain the difference between the position angles
of the blueshifted and redshifted components of the CO($J=3-2$)
outflow (Figure 1).

The accreting disk hypothesis is supported by the detection of the
powerful outflow that emanates with an almost perpendicular
orientation to the object.  The large disk shows to be Keplerian and
with a contracting velocity of about a few km s$^{-1}$ as revealed by
\citet[][]{zapataetal2009}.  The models presented in Figure 4 were
obtained using a slightly different P.A. = 30$^\circ$, inclination
angle of {\it i}=25$^\circ$, a smaller size (3000 AU), and without an
inner cavity than in \citet[][]{zapataetal2009}.

In \citet[][]{zapataetal2009} and here, we have modeled a contracting
flattened disk in Keplerian rotation with a central hole, using the
disk parametrization from Guilloteau, Dutrey \& Simon (1999).  The
contraction is assumed to have the functional form of free-fall
(i.e. V$_{inf}$ $\propto$ $\frac{1}{\sqrt r}$), with a reference
velocity at the reference radius. However, our resolution is
insufficient to determine the exact functional form.  The model is for
the same molecules and transitions.  A better fit than all other
trials in our recurrence was found until we obtained similar
structures in our model to those imaged (Figure 4).  Most of the
physical parameters in the model were constrained in the process. The
model fits the observations reasonably well.

In particular, the molecules HCOCH$_2$OH(62$_{13,49}$-62$_{12,50}$)
and HC$_3$N(38-37) ($\nu_t$=0) revealed the possible presence of a
warm companion located to the northeast of the disk (Figure 2).  We
suggest that the companion could be a consequence of disk
fragmentation due gravitational instabilities because the disk is
extremely large, see for a reference of this phenomenon
\citet{kra2006,kru2009}.
 
\subsection{A possible sequence on the evolution of nascent massive stars?}

The multilayer disk imaged in this paper reveals a possible link
between the Hyper/Ultra-compact HII regions and the hot cores.  These
observations suggest a possible sequence on the formation of massive
stars. We present this simple sequence in Figure 7, and discuss it as
follows:

 \begin{itemize}

\item 
{\it Phase I:} In the first phase, a large and massive pseudo-disk is
formed together with a bipolar outflow from a large and dense core.
The pseudo-disk is surrounded by an infalling envelope with accretion
rates on the order of 10$^{-3}$ M$_\odot$ yr$^{-1}$.  The molecular
emission arising from the circumstellar pseudo-disk could be observed
as a single structure without a well defined temperature gradient
across it. The pseudo-disks observed toward these objects are not
classical cicumstellar disks associated with low-mass stars which are
centrifugally supported, rather these might be contracting
pseudo-disks \citep[see for example the large and massive contracting
  disks associated with the high-mass protostars AFGL490 and NGC
  7538S:][] {naka1991,san2010}. The pseudo-disk could be also likely
circumbinary or circum-multiple because of the large multiplicity
presented on the massive stars.

\item 
{\it Phase II:} The large and massive pseudo-disk formed earlier is
observed now in molecular shells or layers. This object is still
surrounded by an infalling envelope.  A clear temperature gradient is
observed across the disk.  The inner cavity could be formed by
photodissociation due to the high temperatures close to massive stars
or maybe to opacity effects.  In this phase, no free-free emission
from a HCHII region is detected probably because the infalling
molecular gas partially quenches its formation or maybe because the
massive central protostar(s) are/is in an early phase with not high
enough temperatures to ionize their/its surroundings. W51 North is
found here. Another possibility is that maybe there is cluster of late
B-type in middle able of forming a strong multilayer molecular disk,
but not for developing HCHII regions.

\item 
{\it Phase III:} A HCHII region appears in the middle of the large
multilayer pseudo-disk. At this stage the central massive protostar
has the sufficient high temperatures ($>$ 10$^4$ K) to ionize its
surroundings, see for example the cases of NGC7538 IRS1
\citep{fra2004,San2009} and MWC349 \citep{tafo2004}. These sources are
maybe variable on time because of the strong accretion from the large
disk continues toward the massive central star, see also G24.78+0.08
A1 \citep{gal2008}. There are some sources that might be situated at
this phase, see for example, the molecular rotating toroids observed
around the UCHII regions G24.78+0.08 A1 and G5.89-0.39
\citep{su2009,Bel2006}. Moreover, \citet{kla2009} found in a group of
HCHII and UCHII regions rotating molecular structures surrounding
them. Therefore, the molecular toroids or rings around an UCHII are
likely the same structures as in Phase II, but just hotter.  The holes
observed in these toroids might be due to the photodissociation of
some molecules, or maybe to the formation of cavities by the presence
of a multiple system of massive stars as mentioned before. In this
phase the masive young star(s) might be still accreting ionized
material as proposed and described by \citet{Ke2006}.

\item 
{\it Phase IV:} In this phase the HCHII region expands and ionizes the
remnant molecular disk or toroid, and creates an extended H II region
surrounding one or maybe multiple young massive stars. The sequence
from UCHII to extended HII regions has been discussed by
\citet{gar2004,fra2007}. If the central star(s) is/are still accreting
material, probably ionized material, an outflow show also be present
even at this stage.

 \end{itemize}

\section{Summary}

We have observed the massive and young protostar W51 North with the
SMA using a high angular resolution and sensitivity at millimeter and
submillimeter wavelengths. We give a summary of the results as
follows:

\begin{itemize}
\item
We report the presence of a large and single molecular disk around the
object W51 north, and confirm the existence of a single powerful
outflow emanating from the disk;

\item
The molecular emission from the large disk is observed in layers with
transitions characterized by high excitation temperatures in the lower
states (up to 1512 K) being concentrated closer to the central massive
protostar. The molecular emission at low or moderate excitation
temperatures additionally exhibits a central cavity with an angular
size around 0.7$''$.  This multilayer disk connects the molecular ring
and dusty compact disk reported recently toward this object
\citep[][]{zapataetal2009}. The LTE modeling presented in this study
also confirms that the hot gas is only found in the innermost disk.
 
\item
A detail study of the kinematics of the molecular gas together with a
model of a circumstellar disk in Local Thermodynamic Equilibrium shows
that the disk is Keplerian, and is accreting fresh gas to the
protostar possibly quenching the formation of a HCHII region;

\item
The thermal emission of the HCOCH$_2$OH reveals the possible presence
of a warm companion located to the northeast of the disk, perhaps
produced by disk fragmentation or maybe is the interaction of the
outflow a high density zone.  The molecular emission of the SO and
SO$_2$ is observed in the circumstellar disk as well as in the
outflow;

\item
We modeled all lines with a LTE synthetic spectra.  For an assumed
source size of 0.5 - 1 arcsec, our modeling yields a column density of
2 - 6 $\times$ 10$^{16-18}$ cm$^{-2}$, a temperature of 250 - 800 K,
and a linewidth of $\sim$ 5 - 10 km s$^{-1}$.

\end{itemize}

We thank the anonymous referee for many valuable suggestions.  This
research has made extensive use of the SIMBAD database, operated at
CDS, Strasbourg, France, and NASA's Astrophysics Data System.  This
research made use of the myXCLASS program
(https://www.astro.uni-koeln.de/projects/schilke/XCLASS), which
accesses the CDMS (http://www.cdms.de) and JPL
(http://spec.jpl.nasa.gov) molecular data bases.

\bibliographystyle{aa} \bibliography{biblio}

\end{document}